\documentclass[journal]{IEEEtran}
\usepackage{times}
\usepackage{graphicx}
\usepackage{subfigure}
\usepackage{amsmath}
\usepackage{amssymb}
\usepackage{amsbsy}
\usepackage{bm}
\usepackage{float}
\usepackage{cite}
\usepackage{algorithmic}
\usepackage{courier}
\usepackage{xcolor}
\usepackage{listings}
\usepackage[hyphens]{url}
\expandafter\def\expandafter\UrlBreaks\expandafter{\UrlBreaks
      \do\a\do\b\do\c\do\d\do\e\do\f\do\g\do\h\do\i\do\j%
      \do\k\do\l\do\m\do\n\do\o\do\p\do\q\do\r\do\s\do\t%
      \do\u\do\v\do\w\do\x\do\y\do\z\do\A\do\B\do\C\do\D%
      \do\E\do\F\do\G\do\H\do\I\do\J\do\K\do\L\do\M\do\N%
      \do\O\do\P\do\Q\do\R\do\S\do\T\do\U\do\V\do\W\do\X%
      \do\Y\do\Z}
\usepackage{mathptmx}
\usepackage{helvet}
\usepackage{type1cm}
\usepackage{makeidx}
\usepackage{multicol}
\usepackage[bottom]{footmisc}
\newcommand{\me}{\mathrm{e}}

\newcommand{\md}{\,\mathrm{d}}
\newcommand{\defas}{\triangleq}

\newcommand{\V}[1]{\ensuremath{\boldsymbol{#1}}}
\newcommand{\M}[1]{\ensuremath{\boldsymbol{#1}}}
\newcommand{\VT}[1]{\ensuremath{\boldsymbol{#1}}^{\textrm{T}}}

\newcommand{\T}{\ensuremath{}^{\textrm{T}}}

\renewcommand{\P}[1]{\ensuremath{\mathsf{#1}}}

\begin{document}
\lstset{language=Matlab,basicstyle=\small\ttfamily,
commentstyle=\color[rgb]{.133,.545,.133}\small\ttfamily,
keywordstyle=\color[rgb]{0,0,1}\ttfamily\normalsize,
stringstyle=\color[rgb]{.627,.126,.941}\ttfamily\small,
showstringspaces=false,
xleftmargin=0.8cm,resetmargins=true,escapechar=\$,captionpos=b,
breaklines=true,breakatwhitespace=false,breakindent=2cm
,numbers=left,numbersep=2ex,numberstyle=\small\ttfamily,
columns=flexible}
\title{Modelling Periodic Measurement Data\\ Having a Piecewise Polynomial Trend Using the\\ Method of Variable Projection}

\author{Johannes~Handler,~Dimitar~Ninevski~and~Paul~O'Leary
\thanks{This work was partially funded by the COMET program within the K2 Center “Integrated Computational Material, Process and Product Engineering (IC-MPPE)” (Project No 859480). This program is supported by the Austrian Federal Ministries for Transport, Innovation and Technology (BMVIT) and for Digital and Economic Affairs (BMDW), represented by the Austrian research funding association (FFG), and the federal states of Styria, Upper Austria and Tyrol.}%
\thanks{Johannes~Handler,~Dimitar~Ninevski~and~Paul~O'Leary are with the Chair of Automation, Department of Product Engineering, University of Leoben, A8700 Leoben, Austria (email: automation@unileoben.ac.at)}}%
\maketitle
\begin{abstract}
This paper presents a new method for modelling periodic signals having an aperiodic trend, using the method of variable projection. It is a major extension to the IEEE-standard 1057 by permitting the background to be time varying; additionally, any number of harmonics of the periodic portion can be modelled. This paper focuses on using B-Splines to implement a piecewise polynomial model for the aperiodic portion of the signal. A thorough algebraic derivation of the method is presented, as well as a comparison to using global polynomial approximation. It is proven that B-Splines work better for modelling a more complicated aperiodic portion when compared to higher order polynomials. Furthermore, the piecewise polynomial model is capable of modelling the local signal variations produced by the interaction of a control system with a process in industrial applications. The method of variable projection reduces the problem to a one-dimensional nonlinear optimization, combined with a linear least-squares computation. An added benefit of using the method of variable projection is the possibility to calculate the covariances of the linear coefficients of the model, enabling the calculation of confidence and prediction intervals. The method is tested on both real measurement data acquired in industrial processes, as well as synthetic data. The method shows promising results for the precise characterization of periodic signals embedded in highly complex aperiodic backgrounds. Finally, snippets of the m-code are provided, together with a toolbox for B-Splines, which permit the implementation of the complete computation.
\end{abstract}

\section{Introduction}
A portion of this paper was originally published at the I${}^2$MTC conference in 2021~\cite{OLeary2021}. There, the authors showed the advantages of applying the \emph{method of variable projection}~\cite{Golub1973,Golub2003} to identifying the four-parameter model for a sine wave, see the IEEE-standard 1057~\cite{IEEE1057} for an exact definition of the model. This model is commonly used in waveform digitizers, e.g. digital oscilloscopes, analog to digital converters etc~\cite{Hejn2003,Baccigalupi2017}. However,
the authors previous work was primarily motivated by the widespread availability  of low-cost MEMS accelerometers and gyroscopes~\cite{Albarbar2008}; which, commonly create such signals. The ubiquitous applicability of these sensors~~\cite{Lee2012,Wang2017,PREETI2019} calls forth the need for stable and computationally efficient methods.

The four-parameter model assumes a constant background term $d$. This paper extends the method to a much wider class of background signals; that is, the background $d(t)$ is a function of time,
\begin{equation}\label{eqn.basicModel}
	y_m(t) = a\,\sin(\omega t + \phi) + d(t).
\end{equation}
The aim is to make the new methods much more widely applicable in the instrumentation of industrial processes.

A periodic signal, without an integer number of cycles in the measurement period $T$, combined with an aperiodic background $d(t)$, is also the fundamental issue behind spectral leakage when computing Fourier spectra~\cite{Jerri1998}.  Consequently, the methods presented here are, not only relevant to industrial instrumentation, but also relevant to many cases of signal processing. Traditionally, this issue has been addressed using windowing techniques, see the seminal paper by Harris~\cite{harris1978}.

\subsection{Difficulty of separation}
\begin{figure}[]
	\centering
	\includegraphics[width=8cm]{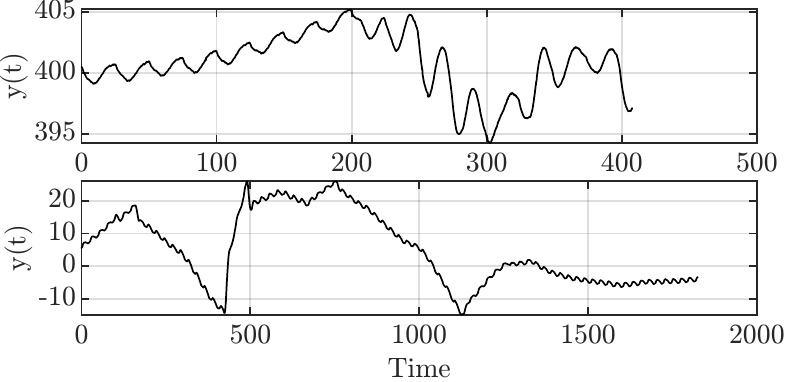}
	\caption{Two real sensor signals measured in different industrial processes. Note there is a global periodic portion to the signal on top of an aperiodic background. The aperiodic portion is due to the process control making changes to the actuators driving the process.}
	\label{fig:IntroSignals}
\end{figure}
In Figure~\ref{fig:IntroSignals}, two real sensor signals measured in different industrial processes\footnote{For confidentiality reasons we are not permitted to name the process behind these measurements.} are shown. Both exhibit a combination of periodic and aperiodic portions. The measurement data is typical for many industrial applications where there is a control process in the background making changes to the actuators driving the process. The goal here is to separate these two components reliably and characterize the periodic portion exactly.

\begin{figure}[]
	\centering
	\includegraphics[width=7.5cm]{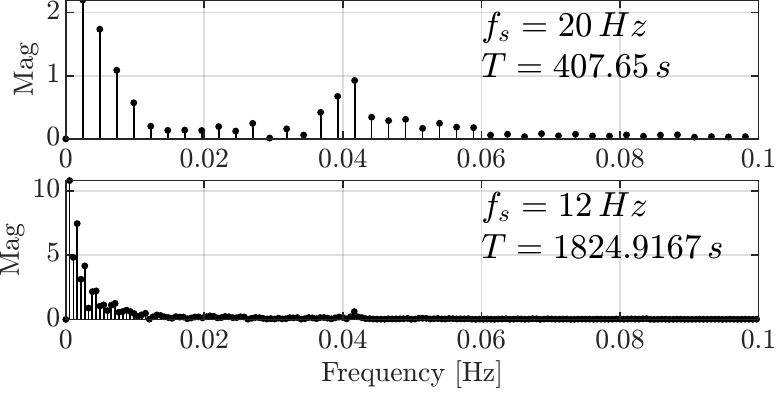}
	\caption{The magnitude of the FFT for the two signals shown in Figure~\ref{fig:IntroSignals}. The frequency range has been cropped to show the region where the signals are to be observed and the DC component removed. The respective sampling frequencies $f_s$ and measurement periods $T$ are also shown.}
	\label{fig:IntroSignalsFFT}
\end{figure}

The magnitude of the FFT for the two sensor can be seen in Figure~\ref{fig:IntroSignalsFFT}. Note: The frequency range has been cropped to show the region where the signals are to be observed. The FFT is highly attractive due to its $\mathcal{C}(n) = n \, log( n )$ computational efficiency~\cite{Brigham1988}. However, as can be seen in Figure~\ref{fig:IntroSignalsFFT}, the spectral leakage from the non-integer number of cycles and from the background signal make the reliable characterization of the periodic portion intractable. Additionally, the sharp changes in the signal, due to the control process, may lead to a Gibbs error with periodic portions.
The problem remains the identification of a suitable model for the $d(t)$ term in Equation~\ref{eqn.basicModel} and then the implementation of a stable, efficient and reliable method to determine the coefficients that lead to a good approximation of the signal.

Similar problems can be found in literature: that is, problems where the model equation is a linear combination of nonlinear functions appear frequently and are commonly solved using the variable projection method. Such problems can be found in power system modal identification \cite{Borden2014}, GPS positioning \cite{ChenGPS2018, Abel1994}, computer vision \cite{Robinson2007, Sheng2008, Aravkin2012}, calibration of measurement data for electromagnetic data inversion \cite{Gao2014}, etc. Many of these problems have a combination of exponential and trigonometric functions as nonlinear basis functions, such as \cite{Borden2014, ChenRegularized2019}. Others \cite{Abel1994, ChenGPS2018, Handler2021} have looked at the case when there are some linear functions, e.g. polynomials, in the nonlinear basis matrix and how to effectively use this to further simplify the calculation. In \cite{Handler2021}, the authors use a combination of polynomials and trigonometric functions to separate the periodic and trend portions of the signal; however, issues arise when the background signal $d(t)$ becomes more complicated and higher degree polynomials are required. Later in this paper it is proven that higher order polynomials, for the background, also start to model portions of the periodic function: This can be seen from the Taylor expansion. Consequently, higher order polynomials can lead to a less accurate characterization of the frequency and magnitude of the periodic portion.

The main contributions  of this paper are:
\begin{enumerate}
	\item To propose a piecewise polynomial as a model for the background signal $d(t)$. This model is compatible and consistent with industrial measurements where the process is being modified by the intervention of actuators. The process control is locally, but not globally, piecewise stationary. This property can be utilized to obtain a better severation from the global periodic oscillation we wish to characterize.
	\item The B-Spline implementation of the piecewise polynomials, permits a direct combination with periodic basis functions; in this manner the method of variable projection can be applied. This splits the originally $n$-dimensional optimization into a one-dimensional nonlinear minimization to determine the frequency $\omega$, followed by an $n-1$ dimensional linear problem given $\omega$. For the linear portion the methods, by definition, ensure that the global minimum given $\omega$ is found. Additionally, the covariance of the linear coefficients is simply computed.
	\item A thorough analysis of the theoretical background to the models
	used and the methods applied to obtain the model coefficients is presented.
	\item Snippets of m-code are provided which implement the proposed methods. An m-code library for the implementation of the B-Spline bases is provided.
	\item Results are presented which validate the method using synthetic data sets with known properties.
	\item The results from the successful application of the new approach to the analysis of industrial measurement data are presented, from two differing processes.
\end{enumerate}

\section{Signal models}

The goal now is to extend the four-parameter model~\cite{IEEE1057} to include a time dependent background $d(t)$. This will accommodate more complex signals, while maintaining the ability to accurately characterize the periodic portion.

\subsection{Piecewise polynomial background}

In this paper we propose a piecewise polynomial~\cite{Wahba1990} for the background signal $d(t)$, implemented via a B-Spline model~\cite[Chapter 2]{PiegTill1996}. The justification for this is: the control signals, which are causal for the aperiodic portion, are in general piecewise $C^0$ continuous. However, due to the dynamics of the process, the observed sensor signals tend to be $C^1$ or $C^2$ continuous. It is the differential equation describing the dynamics of the process, that govern the order $n$ of the continuity $C^n$ observed in the sensor signals. The piecewise $C^n$ continuity of the background signal ensure that piecewise polynomials are suitable models for $d(t)$.
A further advantage in industrial measurements is that the break points, required for the piecewise polynomials, are known from the control signals.

The B-Spline implementation has been chosen, over the pp-spline, since it permits the direct combination of the basis functions $\M{B}_s$ for the spline and $\M{B}_p$ for the periodic portions, into one consistent set of matrix equations. In this manner the method of variable projection~\cite{Golub1973,Golub2003} can be applied, during the optimization, to separate the nonlinear from the linear portion of the computation. This improves the stability of the optimization process and reduces the computational complexity. The possibility of a piecewise continuous background $d(t)$, extends greatly the areas where this methods can be applied in the analysis of measurement signals.
The four parameter model~\cite{IEEE1057}, considered in the previous paper~\cite{OLeary2021}, is now a special case also covered by the model proposed here.

\subsection{B-Spline notation}

To compute a complete set of spline basis functions we need to know the vector of points $\V{x}$, where the spline is to be evaluated, the desired degree of the spline $\delta$ and a vector of knots $\V{\kappa}$. Consequently, the notation\footnote{This notation is compatible with the B-Spline m-code implementation made available at MATLAB-FileExchange~\cite{FileExchange}.} for the spline should be of the form $\M{B_s}(\V{x}, \delta, \V{\kappa})$. The spline portion of the model $\V{y}_s$ is now computed as a linear combination of these basis functions with the coefficient vector $\V{\beta}$, i.e.,
\begin{equation}
	\V{y}_s = \M{B_s}(\V{x}, \delta, \V{\kappa}) \,\, \V{\beta}.
\end{equation}
The number of basis functions required and with this the length of the coefficient vector $\dim(\beta)$ is dependent of $\delta$ and the number of knots, i.e.,
\begin{equation}\label{eqn.dimBeta}
	\dim(\beta) = \dim(\V{\kappa}) + \delta - 1
\end{equation}

\subsection{Periodic notation}
The periodic portion is modelled as a base frequency $\omega$, together with a number of harmonics $\nu$ that need to be considered. The use of harmonics permits the modelling of more complex periodic signals. The complete notation required for the periodic basis functions is $\M{B_p}(\V{x}, \omega, \nu)$. The coefficient vector $\alpha$ has two components per harmonic,
\begin{equation}\label{eqn.dimAlpha}
	\dim(\V{\alpha}) = 2 \, \nu,
\end{equation}
since one sine and one cosine is required per harmonic. The periodic portion of the signal is computed as,
\begin{equation}
	\V{y}_p = \M{B_p}(\V{x}, \omega, \nu) \,\, \V{\alpha}.
\end{equation}
\subsection{Complete signal model}
The complete signal model $\V{y}_m = \V{y}_p + \V{y}_s$, which in matrix form can now be written as,
\begin{equation}
	\V{y}_m =
	\begin{bmatrix}
		\M{B_p}(\V{x}, \omega, \nu) & \M{B_s}(\V{x}, \delta, \V{\kappa})
	\end{bmatrix}
	\,\,
	\begin{bmatrix}
		\V{\alpha} \\
		\V{\beta}
	\end{bmatrix}.
\end{equation}
To obtain a less cumbersome notation we shall define some simplified notations: since the locations of the samples $\V{x}$, $\nu$ the number of harmonics, $\delta$ the degree of the polynomial spline and $\V{\kappa}$ the vector of break points do not change during the optimization, we shall define
\begin{equation}
	\M{B}(\omega) \defas
	\begin{bmatrix}
		\M{B_p}(\V{x}, \omega, \nu) & \M{B_s}(\V{x}, \delta, \V{\kappa})
	\end{bmatrix}
	\,\,\text{given}\,\, \V{x},\nu,\delta,\V{\kappa},
\end{equation}
indicating that $\M{B}(\omega)$ is a matrix of basis functions nonlinear in $\omega$ and $\omega$ is relaxant during the optimization.  Additionally,
\begin{equation}
	\V{\gamma} \defas
	\begin{bmatrix}
		\V{\alpha} \\
		\V{\beta}
	\end{bmatrix}.
\end{equation}
Now, the notation for the model calculation simplifies to,
\begin{equation}
	\V{y}_m = \M{B}(\omega) \, \, \V{\gamma}.
\end{equation}
The least squares approximation of the signal $\V{y}$ by the model $\V{y}_m$, now requires us to minimize,
\begin{equation}\label{eqn.CostFcn}
	\min_{\omega, \V{\gamma}} \| \V{y} - \M{B}(\omega) \, \V{\gamma}\|^2_2.
\end{equation}
Attacking this task directly leads to a very high dimensional nonlinear optimization problem, with all the associated numerical difficulties. Fortunately, the problem separates into a linear combination of basis functions which are nonlinear in $\omega$. This makes the system of equations suitable for the application of variable projections.

\section{Method of Variable Projection}
\label{sec: VarProMethod}
The method of variable projection~\cite{Golub1973,Golub2003} is a method for solving separable nonlinear least squares problems. It is characterized by the model being a linear combination of nonlinear functions, meaning
\begin{equation}
	\label{eq: VarProModelEq}
	y_m = \alpha_1b_1\left(\V{\delta}\right) + \alpha_2 b_2\left(\V{\delta}\right) + \hdots + \alpha_k b_k\left(\V{\delta}\right),
\end{equation}
where $\V{\delta}$ is a vector of $j$ nonlinear coefficients, $b_i\left(\V{\delta}\right)$ are the basis functions, which are nonlinear in $\V{\delta}$ and $\alpha_i$ are the $k$ linear coefficients.
Concatenating the individual basis functions into the matrix of basis functions, yields
\begin{equation}
	\M{B}(\V{\delta}) = [\V{b}_1(\V{\delta}), \V{b}_2(\V{\delta}), \hdots,  \V{b}_k(\V{\delta})].
\end{equation}
The notation $\M{B}(\V{\delta})$ indicates that the contents of the matrix $\M{B}$ are dependent on $\V{\delta}$.
Now defining the coefficient vector
\begin{equation}
	\V{\alpha} \defas [\alpha_1, \alpha_2,\hdots \alpha_k]^\mathrm{T}
\end{equation}
leads directly to the matrix vector equation,
\begin{equation}\label{eqn:linsol1}
	\V{y}_m = \M{B}(\V{\delta}) \, \V{\alpha}.
\end{equation}
Given an estimate for $\V{\delta}$, which yields a stationary value for the cost function, a least squares estimate for $\V{\alpha}$ is obtained from,
\begin{equation}\label{eqn:linsol2}
	\V{\alpha} = \M{B}^+(\V{\delta}) \, \V{y},
\end{equation}
whereby, $\M{B}^+(\V{\delta})$ denotes the Moore-Penrose pseudo inverse of $\M{B}(\V{\delta})$. Now substituting Equation~\ref{eqn:linsol2} into~\ref{eqn:linsol1} one obtains,
\begin{equation}
	\V{y}_m = \M{B}(\V{\delta}) \, \M{B}^+(\V{\delta}) \, \V{y}.
\end{equation}
Defining $\M{P}(\V{\delta}) \defas \M{B}(\V{\delta}) \, \M{B}^{+}(\V{\delta})$, this is the projection onto the subspace spanned by the basis functions contained in $\M{B}(\V{\delta})$. This projection varies with $\V{\delta}$, hence the name, method of variable projection. The residual vector is computed as
\begin{equation}
	\V{r} = \V{y} - \V{y}_m
\end{equation}
leading to the cost function $E(\V{\delta})$, defined as the sum of squares of the residual $\V{r}$, to be calculated as
\begin{align}\label{eqn:costFn}
	E(\V{\delta}) &= \| \V{r} \|^2 \\
	&= \| \V{y} - \M{B}(\V{\delta}) \, \M{B}^+(\V{\delta}) \, \V{y} \|^2,\\
	&= \| \left\{\M{I} - \M{B}(\V{\delta}) \, \M{B}^+(\V{\delta})\right\} \, \V{y} \|^2.
\end{align}
This is called the \emph{variable projection functional} (VPF). Note that the estimation of $\V{\delta}$ is now a nonlinear least squares problem in $j$ dimensions; whereas, the original problem has $j + k$ coefficients.
Equation~\ref{eqn:costFn} permits an explicit calculation of the cost function $E(\V{\delta})$ as a function of $\V{\delta}$ for a given measurement $\V{y}$.
In Section~4 of~\cite{Golub1973}, Golub provides the formal proofs required to determine that the Fr\'{e}chet derivatives over the pseudo-inverse and projection yield the gradient of the cost function. This implies that gradient based nonlinear solvers can be used to find the value of $\V{\delta}$ which minimizes the cost function $E(\V{\delta})$.

This method is particularly advantageous for the case at hand, since the cost function is only nonlinear in the variable $\omega$. From Equation~\ref{eqn.CostFcn} we obtain the variable projection (cost) function,
\begin{equation}
	E(\omega) = \| \V{y} - \M{B}(\omega) \, \M{B}^{+}(\omega) \, \V{y} \|^2_2.
\end{equation}
That is, the problem has reduced to a one-dimensional nonlinear optimization problem.
\section{Covariance propagation}
A further advantage of the method of variable projection is that it yields a direct method of calculating the covariance propagation for the linear coefficients. Given $\V{\gamma} = \M{B}^+(\omega) \, \V{y}$, a linear mapping, then $\M{\Lambda}_{\gamma}$, the covariance matrix of $\V{\gamma}$, can be computed as~\cite{Brandt1998},
\begin{equation}
	\M{\Lambda}_{\gamma} = \M{B}(\omega)^+\, \M{\Lambda}_y \, {\M{B}(\omega)^+}^\mathrm{T}.
\end{equation}
whereby, $\M{\Lambda}_y$ is the covariance matrix of the data vector $\V{y}$. If $\V{y}$ is perturbed by i.i.d. Gaussian noise with the standard deviation $\sigma_y$, then the equation becomes
\begin{equation}
	\M{\Lambda}_{\gamma} = \sigma^2_y \, \M{B}(\omega)^+ \, {\M{B}(\omega)^+}^\mathrm{T}.
\end{equation}
Assuming the model is bias free\footnote{In applications the Gaussian nature of the residual vector $\V{r}$ should be verified to ensure that this assumption is valid.}, an estimate for $\sigma_y^2$ can be computed from the residual vector $\V{r}$ as follows:
\begin{equation}
	\sigma_y^2 = \frac{1}{n - n_{df}}\,\| \V{r} \|_2^2,
\end{equation}
where by $n_{df}$ denotes the number of degrees of freedom. In this manner, the covariance of the linear coefficients can be  computed directly.

\section{Interaction between the portions}
The question now is: what would be an ideal signal for $d(t)$?
Ideally the model for the background $d(t)$ would be fully orthogonal to the periodic portion, to ensure there is no interaction between the components. The inner product of two vectors $c = \VT{a} \, \V{b} = |\V{a}|\,|\V{b}|\,\cos(\phi)$, where $\phi$ is the angle between the two vectors, i.e., it is a measure for their orthogonality. The inner product for a continuous $d(t)$ wrt. to a single periodic component $y_p(t) = \me^{- j \omega t}$ can be calculated as,
\begin{equation}
	c = \int_{-\infty}^{\infty} d(t) \, \me^{- j \omega t} \md t.
\end{equation}
This corresponds to the fourier coefficients for $d(t)$ up to a factor of scale. Consequently, for discrete $d(t)$ we can apply the FFT to obtain a numerically efficient estimate for the interaction between the two portions of the signal.

\begin{figure}[]
	\centering
	\includegraphics[width=8cm]{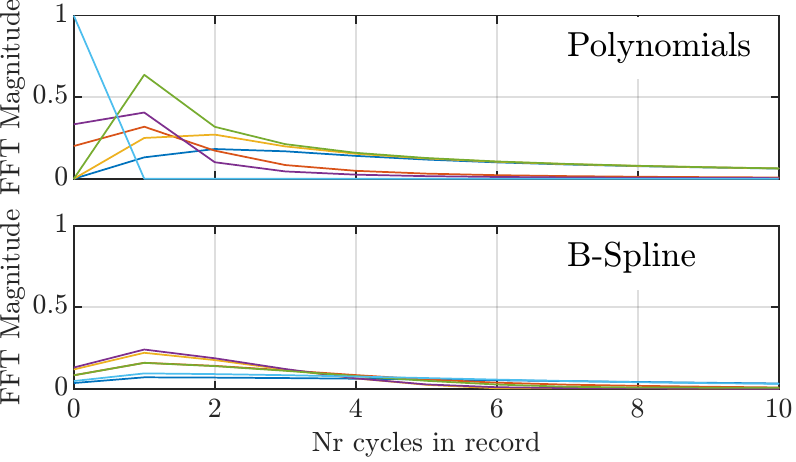}
	\caption{Top: Fourier spectra of the monomials $x^0 \ldots x^5$ in the interval $-1 \leq x \leq 1$. Bottom: The Fourier spectra for the $n=6$ B-Spline basis functions used for the synthetic data, see Section~\ref{subsec.Synthetic}. Note: there is significantly less interaction of the B-Splines with the periodic signal.  }
	\label{fig:InteractionsSpectra}
\end{figure}

The Weierstrass theorem~\cite{Weierstrass1885} states that, in a finite interval, any function can be approximated by a polynomial with the desired accuracy, given a polynomial of sufficient degree. This is also true for periodic signals; consider the Maclaurin series for $\sin(x)$ and $\cos(x)$,
\begin{align}
	\sin \left(x\right) &= x-\frac{x^3}{3!}+\frac{x^5}{5!}-...=
	\sum_{k=0}^{\infty}\left(-1\right)^k\frac{x^{2k+1}}{\left(2k+1\right)!}\\
	\cos\left(x\right) &= 1-\frac{x^2}{2!}+\frac{x^4}{4!}-...=
	\sum_{k=0}^{\infty}\left(-1\right)^k\frac{x^{2k}}{\left(2k\right)!}.
\end{align}
Consequently, a global polynomial, as a model for $d(t)$, will have an interaction with the periodic portion. The level of interaction can be computed by creating the Vandermonde matrix $\M{V}$ for $-1 \leq x \leq 1$, and applying the FFT to its columns. This yields the spectra of the monomials, see Figure~\ref{fig:InteractionsSpectra}.

In the case of B-Splines\cite{DeBoorSplines} the interaction depends on the locations of the knots. Note that when using splines, determining the locations of the breakpoints (knots) is very important and often very challenging. Here, however, the focus is on the variable projection method and for further information on the placement of knots, the reader is referred to \cite{Dung2017, Ninevski2020}. Fortunately, in the application fields driving this work, the locations of the knots are known from the control system data.

At this point the knots associated with the synthetic data, see Section~\ref{subsec.Synthetic}, are used; since this permits a comparison of B-Spline bases with monomials. The interaction between the B-Splines and the periodic signal is also shown in Figure~\ref{fig:InteractionsSpectra}.

The Fourier spectra of the monomials and the B-Spline bases, show that there is significantly less interaction between the periodic signal and the B-Splines. Furthermore, the interaction will diminish with increasing number of knots, since the B-Spline become more local in nature. This is the reason why using polynomials, of higher degree, to model a trend globally does not always provide good results \cite{Handler2021}. Consequently, the B-Splines can be considered as the preferred solution for $d(t)$ over global polynomials.
\section{Numerical testing}
The proposed method was tested on different datasets: a synthetic dataset, and two real sensor signals emanating from an industrial process, where a control process is making changes to the actuators driving the process.
\subsection{Synthetic examples}\label{subsec.Synthetic}
\begin{figure}[]
	\centering
	\includegraphics[width=0.95\columnwidth]{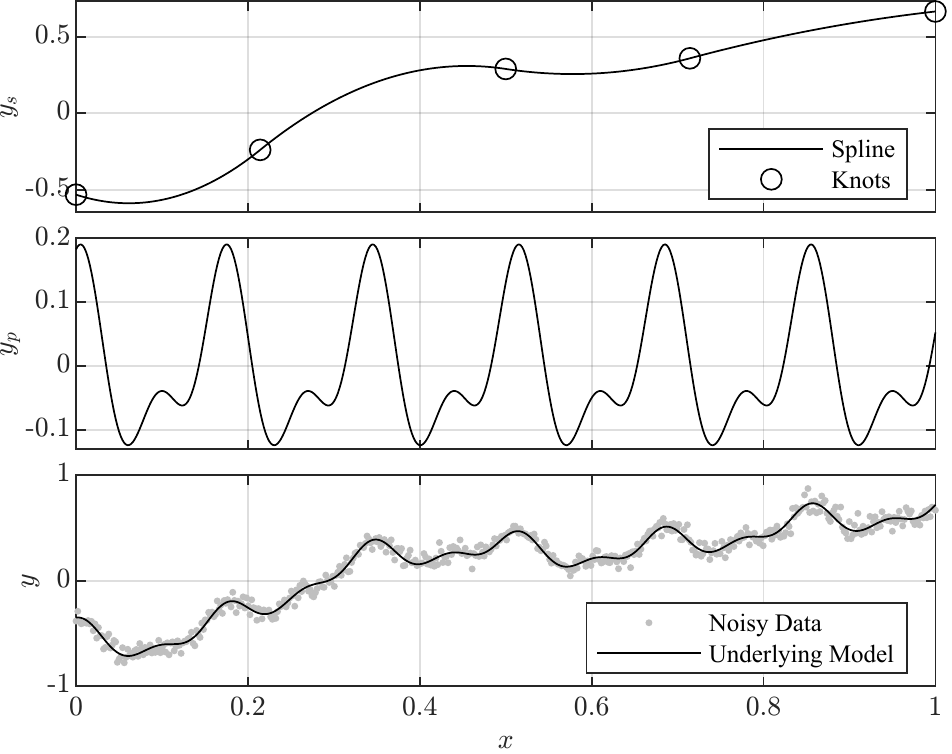}
	\caption{The two components of the synthetic test signal and the final test signal defined for a normalized time vector $x$: (top) the piecewise polynomial trend, defined by 5 knots and with the degree of the polynomial spline being $\delta=2$, (middle) the periodic component consisting of a sine wave with frequency $\omega=36.96$ and its first two harmonics, (bottom) the sum of those two components superimposed with i.i.d. Gaussian noise with a standard deviation of $\sigma= 0.05$. }
	\label{fig:SyntheticData1}
\end{figure}

For the purposes of testing, a synthetic data set was generated as the sum of a spline and a trigonometric function, as shown in Figure~\ref{fig:SyntheticData1}. Here the knot locations $\kappa$ were known a-priori. The piecewise polynomial trend is defined by five knots. The degree of the spline is chosen to be $\delta=2$. The periodic component consists a sine with frequency $\omega=36.96$ and its first two harmonics. Finally, the synthetic signal is the sum of those two components superimposed with i.i.d. Gaussian noise with a standard deviation of $\sigma= 0.05$ (Figure~\ref{fig:SyntheticData1}, bottom).

\begin{figure}[]
	\centering
	\includegraphics[width=0.9\columnwidth]{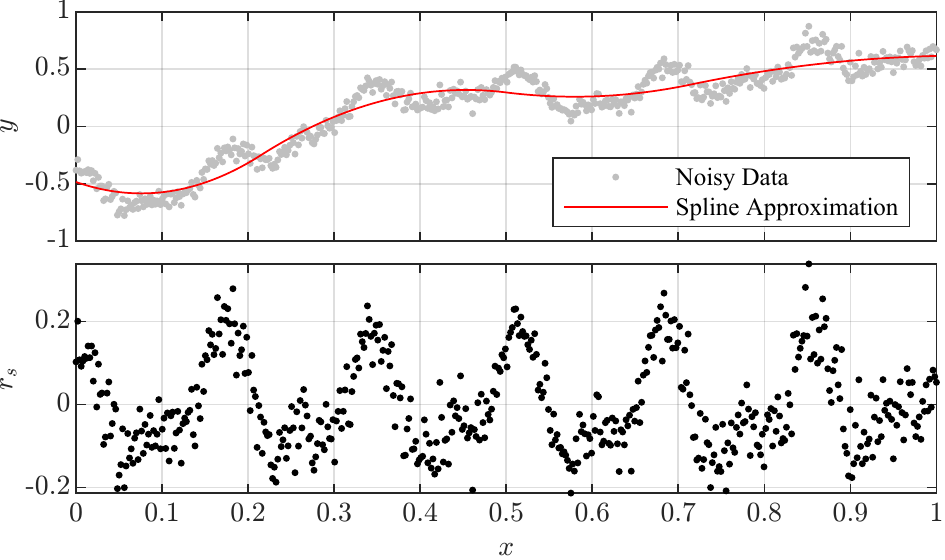}
	\caption{For the computation of a suitable initial value, the measurement data is solely approximated by the B-Spline basis functions: (top)  synthetic data and the approximating B-Spline, (bottom) the residual of the approximation.}
	\label{fig:InitValTimeDom}
\end{figure}

\begin{figure}[]
	\centering
	\includegraphics[width=0.9\columnwidth]{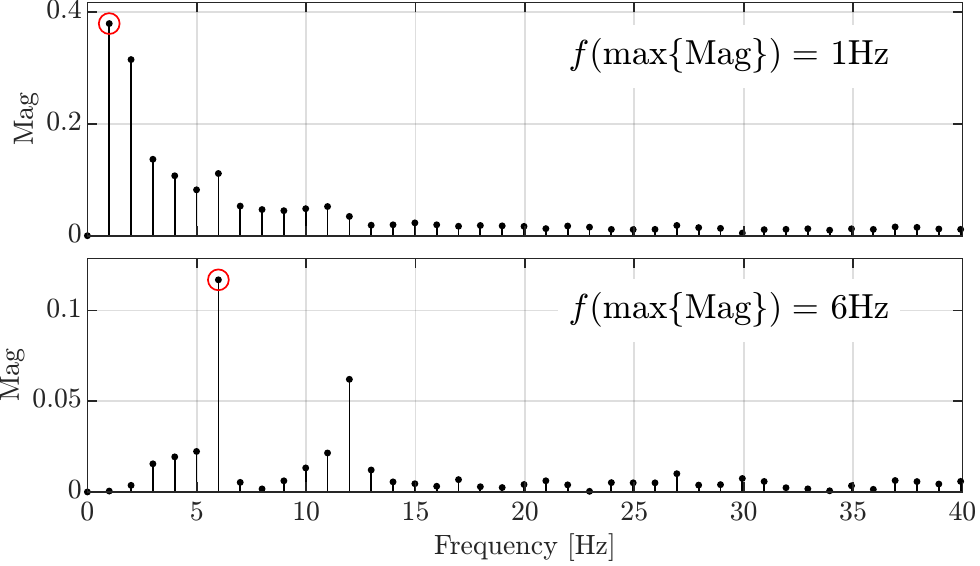}
	\caption{The magnitude of the FFT for the two signals shown in Figure~\ref{fig:InitValTimeDom}: (top) The spectrum of the synthetic data, (bottom) the spectrum of the residual. }
	\label{fig:InitValSpectrum}
\end{figure}

\begin{figure}[]
	\centering
	\includegraphics[width=0.9\columnwidth]{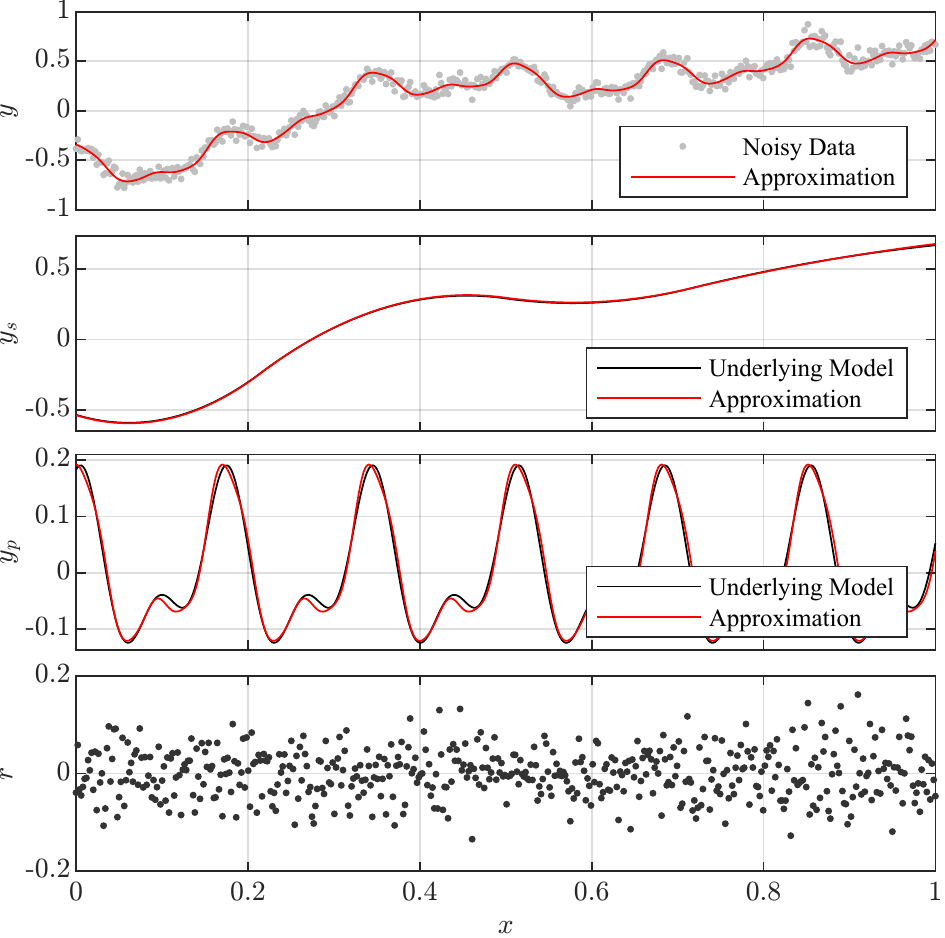}
	\caption{The result of the data approximation by the method of variable projection: (top) The noisy data, and the estimated function, (middle two) the resulting spline and periodic components, (bottom) the residual. }
	\label{fig:SyntheticData2}
\end{figure}

In general, the result of a numerical optimization strongly depends on the initial value. Here, the FFT provides a good starting point for the numerical iteration procedure. However, applying the FFT directly on the measurement signal $y$ does not provide good results; since the aperiodic background $d(t)$ leads to a major distortion of the spectrum, see Figure~\ref{fig:InitValSpectrum} (top). In order to reduce this influence the signal $y$ is first approximated solely by the B-Spline basis functions, 
\begin{equation}
	\V{y}_s = \M{B_s} \, \M{B_s}^+ \, \V{y} .
\end{equation}
The residual of this approximation, $\V{r}_s = \V{y}-\V{y}_s$, contains primary periodic portions, as shown in Figure~\ref{fig:InitValTimeDom}. The frequency corresponding to the maximum magnitude in the spectrum of the residual $r_s$ is the base frequency we are looking for, subject to Gibbs error~\cite{Jerri1998}. However, this value is a good initial value for the nonlinear optimization process to get a more accurate approximation for the base frequency.\\
In the presented example this gives an initial frequency of $\omega_{init}=37.699$ and lead to an estimation of $\hat{\omega}=36.884$.
The final result, as well as the residual, is shown in Figure~\ref{fig:SyntheticData2}. These results demonstrate the ability of the computational method to successfully separate the components of the signal.\\
Additionally, for this example, the covariances of the linear coefficients were calculated. Since a spline of degree $2$ with $5$ knots was used, according to Equation~\ref{eqn.dimBeta} this results in $6$ spline coefficients denoted by $\beta_1, \beta_2, \hdots, \beta_6$. The covariances of those coefficients can be seen in Table~\ref{tab:CovarSplines}. Note that this model has more linear coefficients, namely the linear coefficients in front of the trigonometric functions, however these were not displayed here for conciseness.
\begin{table}[H]
	\begin{center}
		\begin{tabular}{l c c c c c c}
			\hline
			Cov & $\beta_1$& $\beta_2$& $\beta_3$& $\beta_4$& $\beta_5$& $\beta_6$\\
			\hline
			$\beta_1$ & 0.132& -0.053& 0.023& -0.010& 0.009& -0.007\\
			$\beta_2$ & -0.053& 0.091& -0.044& 0.019& -0.015& 0.008\\
			$\beta_3$ & 0.023& -0.044& 0.063& -0.029& 0.022& -0.010\\
			$\beta_4$ & -0.010& 0.019& -0.030& 0.053& -0.041& 0.019\\
			$\beta_5$ & 0.009& -0.015& 0.022& -0.041& 0.100& -0.054\\
			$\beta_6$ & -0.007& 0.008& -0.010& 0.019& -0.054& 0.105 \\
			\hline
		\end{tabular}
	\end{center}
	\caption{Table of covariances for the test case shown in Figure~\ref{fig:SyntheticData2}. The results are scaled by $10^{3}$ since they were very small.}
	\label{tab:CovarSplines}
\end{table}
\subsection{Industrial measurement data}
The presented algorithm was further tested on data sets emanating from different industrial processes.\footnote{Due to confidentiality reasons the original data sets were anonymized.}
A coordinate transformation is applied to $x$ to yield dimensionless values. The x-scaling has been chosen using apriory knowledge, so that the expected periodicity has a period of $T_e=1$. As mentioned before, choosing the right locations of the breakpoints $\kappa$ and the degree of the spline $\delta$ has a strong influence on the quality of the result. Here, the knots are chosen to correspond to time points when the control system influences the process via an actuator. This has the advantage that the knot locations can be derived directly from the control signals.
 
\begin{figure}[]
	\centering
	\includegraphics[width=0.95\columnwidth]{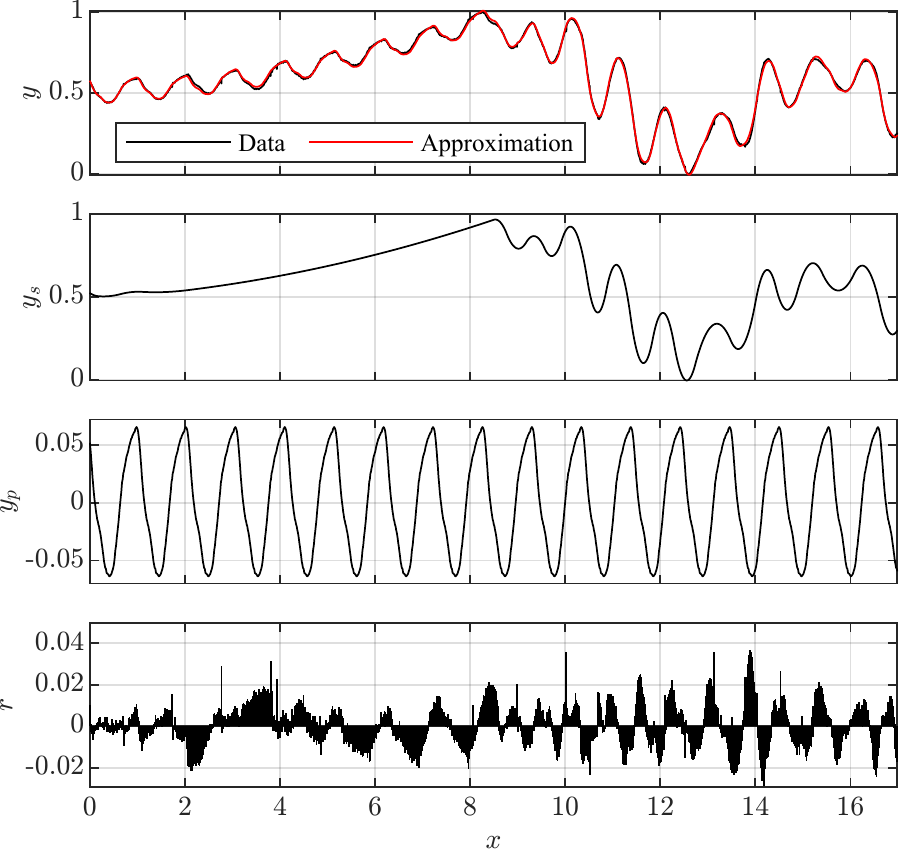}
	\caption{Measurement data from an industrial process. (top) Measurement signal and the approximated function, whereby the period was identified to be $1.04$. (middle two) The spline and periodic portion of the signal obtained using the method of variable projection. (bottom) Residual of the approximation.   }
	\label{fig:RealLifeData1}
\end{figure}

In the first example, shown in Figure~\ref{fig:RealLifeData1}, it can be seen that the aperiodic component $y_s$ exhibits an oscillatory behavior at around $x=8$. This is due to the local activation of the process dynamics by the process control. Due to the local nature of splines they are well suited to model this behavior. The natural frequency of the process dynamics is close to the frequency of the global periodic component $y_p$. Despite this fact the proposed algorithm is still in the position to successfully separate the two signal components. Also the period being identified as $T_i=1.04$ is in the expected range. This clearly speaks for the robustness of the method. 

\begin{figure}[]
	\centering
	\includegraphics[width=0.95\columnwidth]{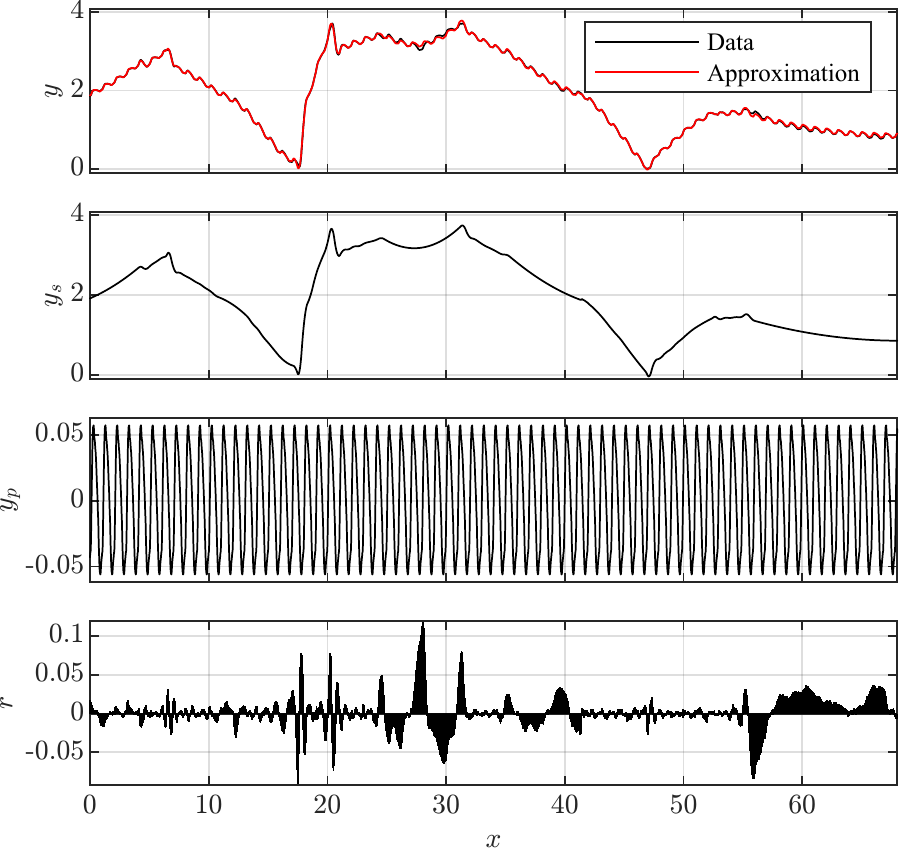}
	\caption{Another industrial measurement data set. (top) Measurement signal and the approximated function, whereby the cycle duration was identified to be $0.996$. (middle two) The spline and periodic portion of the signal obtained using the method of variable projection. (bottom) Residual of the approximation.}
	\label{fig:RealLifeData2}
\end{figure}

Data from a second process is presented in Figure~\ref{fig:RealLifeData2}. Here the resulting period after optimization is $T_i=0.996$. In this measurement data are some areas, e.g. $x=28$ and $x=56$ where the residual is relatively large; this requires further discussion. Considering the residual in more detail: the disturbances are primarily impulse in nature and not periodic; this indicates that $d(t)$ may need to be modified to improve the modelling. The local nature of B-Splines permits the insertion of additional knots~\cite{DeBoorSplines}. The new breakpoints should be inserted where the local integral of $r(t)^2$ is a maximum: This is an area where further research is justified. Furthermore, this would open the possibility to identify and quantify extraneous influences not emanating from the control.

\section{Code implementation}
Here the salient snippets of code required to implement the method of variable projection for modelling a periodic signal having a piecewise polynomial background are presented. The necessary support functions\footnote{Harmonic basis functions are choosen instead of a single sine wave to model more complex periodic signals.} \lstinline{harmonicBasis}and \lstinline{bSplineBasis} are made available at MATLAB-FileExchange~\cite{FileExchange}. The first piece of m-code is a function that is used to compute the cost function $E(\omega)$ for the method of variable projection, see Listing~1.
\begin{lstlisting}[caption={Variable projection cost function},label={lst:cost}]
function cost = varproCost( omega, x, y, nu, delta, kappa )
%
% Setup the basis functions
Bp = harmonicBasis( x, omega, nu );
Bs = bSplineBasis( x, delta, kappa );
B = [ Bp, Bs ];
% Compute the projection
ym = B * ( B \ y );
% Evaluate the residual
r = y - ym;
% Compute the corresponding cost
cost = norm( r );
\end{lstlisting}
As in~\cite{DPOLeary2013}, here a standard high quality generic nonlinear
iteration process\footnote{This is a common approach, since such solvers are available in most numerical computation packages.} is availed to perform the optimization. The code required for this portion is shown in Listing~2. An autonomous function \lstinline{fun} is defined with one calling parameter, \lstinline{omega} and one return parameter \lstinline{cost}. Additionally, there are five passively passed parameters \lstinline{x, y, nu, delta} and \lstinline{kappa}: the time vector and measurement vector, respectively, as well as the parameters defining the B-Spline bases and the number of harmonics.
\begin{lstlisting}[caption={Wrapping with a nonlinear solver}]
% Define the anonymous for the cost
fun = @(omega) varproCost( omega, x, y, nu, delta, kappa );
% Wrap with the nonlinear iteration.
omega = lsqnonlin( fun, omega_Init );
\end{lstlisting}
At the end of running the code in Listing~2, the optimal value for $\omega$ is obtained. The remaining linear coefficients $\V{\gamma}$, since $\omega$ is available, can be computed according to Equation~\ref{eqn:linsol2}, this is implemented in Listing~3.
\begin{lstlisting}[caption={Compute the linear parameters.},label={lst:cost}]
Bp = harmonicBasis( x, omega, nu );
Bs = bSplineBasis( x, delta, kappa );
B = [ Bp, Bs ];
% Compute the projection
cfs = B \ y ;
\end{lstlisting}
Finally, if required the covariances of the linear coefficients can be computed according to the m-code in Listing~4.
\begin{lstlisting}[caption={Code required to compute the covariances.}]
  nrParams = size( B, 2 ) + 1;
  % Number of degrees of freedom
  df = length(y) - nrParams ;
  % Estimate the standard deviation of y
  stdY = norm(r) / sqrt(df);
  % Evaluate the covariance
  Bp = pinv( B );
  Cov = stdY^2  * ( Bp * Bp' ) ;
\end{lstlisting}
When considering the cases with other basis functions for the aperiodic component, e.g. exponentials~\cite{Borden2014, ChenRegularized2019} or polynomials~\cite{Handler2021}, only line 5 in Listing~1 and line 2 in Listing~3 need to be adjusted, according to the appropriate basis functions. The rest of the code remains the same.

\section{Further extension}
The approach described here could be further generalized to approximate functions using rational polynomials, i.e. functions of the form
\begin{equation}
	r\left(x, \alpha, \beta\right) = \frac{f\left(x, \alpha\right)}{g\left(x, \beta\right)}.
\end{equation}
Such functions are commonly encountered in IIR-Filter design, in that case polynomials in $z$. Zolotarev~\cite{Istace1995} found solutions to two specific cases based on elliptical functions, but no general solutions. There are also solutions based on Yule-Walker methods~\cite{Friedlander1984}. However, none of these previous solutions take advantage of the method of variable projection to reduce the dimensionality of the nonlinear optimization.

Let $\M{B}_f(\V{x}, \V{\alpha}), \M{B}_g(\V{x}, \V{\beta})$ be  polynomial basis matrices, such that the polynomials can be discretely written as
\begin{equation}
	f\left(x, \V{\alpha}\right) = \M{B}_f(\V{x}, \V{\alpha}) \V{\alpha}, \quad g\left(x, \V{\beta}\right) = \M{B}_g(\V{x}, \V{\beta}) \V{\beta}.
\end{equation}
Then one could write
\begin{equation}
	r\left(x, \V{\alpha}, \V{\beta}\right) = \underbrace{\left\{\M{B}_g(\V{x}, \V{\beta}) \V{\beta}\right\}^- \M{B}_f(\V{x}, \V{\alpha})}_{\M{M}(\V{x}, \V{\beta})} \V{\alpha} = \M{M}(\V{x}, \V{\beta})\V{\alpha}
\end{equation}
where $\left\{\M{B}_g \V{\beta}\right\}^-$ denotes some generalized inverse, which needs to be further investigated.
So if the function which needed to be approximated was $d\left(x\right)$ and its shape was similar to the shape of a rational function, one would write
\begin{equation}
	d\left(x\right) \approx \M{M}(\V{x}, \V{\beta})\V{\alpha}
\end{equation}
which is a linear combination of nonlinear basis functions contained in the matrix $\M{M}(\V{x}, \V{\beta})$, and thus the method of variable projection could be used.
\section{Conclusions}
It has been shown that the new method permits the characterization of periodic portions of a signal embedded in complex aperiodic backgrounds, cases where the IEEE-standard 1057 is inappropriate. The use of $C^n$ continuous B-Splines for the background is compatible with the piecewise local interactions of a control system with an industrial process. It can be concluded that the new method is a contribution to improving the instrumentation of industrial processes. This new implementation works especially well when the periodic portion is not perfectly periodic in its observation time, since it avoids the Gibbs error and spectral leakage associated with classical Fourier techniques. It also avoids the undesirable effects of having to apply windowing to limit leakage. The algebraic formulation, also yielded a simple approach to calculating the covariance of the linear coefficients, i.e., all but one of the coefficients. Consequently, the method provides a computation means of dealing with uncertainty. The m-code snippets show the simplicity of the coding required to implement the complete analysis. 

\bibliographystyle{IEEETran}
\bibliography{sineFitV2,newCitations}

\begin{thebibliography}{10}
\providecommand{\url}[1]{#1}
\csname url@samestyle\endcsname
\providecommand{\newblock}{\relax}
\providecommand{\bibinfo}[2]{#2}
\providecommand{\BIBentrySTDinterwordspacing}{\spaceskip=0pt\relax}
\providecommand{\BIBentryALTinterwordstretchfactor}{4}
\providecommand{\BIBentryALTinterwordspacing}{\spaceskip=\fontdimen2\font plus
\BIBentryALTinterwordstretchfactor\fontdimen3\font minus
  \fontdimen4\font\relax}
\providecommand{\BIBforeignlanguage}[2]{{%
\expandafter\ifx\csname l@#1\endcsname\relax
\typeout{** WARNING: IEEEtran.bst: No hyphenation pattern has been}%
\typeout{** loaded for the language `#1'. Using the pattern for}%
\typeout{** the default language instead.}%
\else
\language=\csname l@#1\endcsname
\fi
#2}}
\providecommand{\BIBdecl}{\relax}
\BIBdecl

\bibitem{OLeary2021}
P.~O'Leary and D.~Ninevski, ``Estimating parameters of a sine wave by the
  method of variable projection,'' in \emph{2021 IEEE International
  Instrumentation and Measurement Technology Conference (I2MTC)}, 2021, pp.
  1--6.

\bibitem{Golub1973}
G.~H. Golub and V.~Pereyra, ``The differentiation of pseudo-inverses and
  nonlinear least squares problems whose variables separate,'' \emph{SIAM
  Journal on Numerical Analysis}, vol.~10, no.~2, pp. 413--432, 1973.

\bibitem{Golub2003}
G.~Golub and V.~Pereyra, ``Separable nonlinear least squares: the variable
  projection method and its applications,'' \emph{Inverse Problems}, vol.~19,
  pp. R1--R26(1), 01 2003.

\bibitem{IEEE1057}
IEEE, ``\uppercase{IEEE} standard for digitizing waveform recorders,''
  \emph{IEEE Std 1057-2017 (Revision of IEEE Std 1057-2007)}, pp. 1--0, 2018.

\bibitem{Hejn2003}
K.~{Hejn} and A.~{Pacut}, ``Effective resolution of analog to digital
  converters,'' \emph{IEEE Instrumentation Measurement Magazine}, vol.~6,
  no.~3, pp. 48--55, 2003.

\bibitem{Baccigalupi2017}
A.~{Baccigalupi}, M.~{D’Arco}, and A.~{Liccardo}, ``Parameters and methods
  for adcs testing compliant with the guide to the expression of uncertainty in
  measurements,'' \emph{IEEE Transactions on Instrumentation and Measurement},
  vol.~66, no.~3, pp. 424--431, 2017.

\bibitem{Albarbar2008}
A.~{Albarbar}, S.~{Mekid}, A.~{Starr}, and R.~{Pietruszkiewicz}, ``Suitability
  of mems accelerometers for condition monitoring: An experimental study,''
  \emph{Sensors}, vol.~8, no.~2, pp. 2192--2196, 2008.

\bibitem{Lee2012}
J.~S. {Lee}, S.~{Choi}, S.~{Kim}, C.~{Park}, and Y.~G. {Kim}, ``A mixed
  filtering approach for track condition monitoring using accelerometers on the
  axle box and bogie,'' \emph{\uppercase{IEEE} Transactions on Instrumentation
  and Measurement}, vol.~61, no.~3, pp. 749--758, 2012.

\bibitem{Wang2017}
H.~{Wang}, Z.~{Liu}, A.~{Núñez}, and R.~{Dollevoet}, ``Identification of the
  catenary structure wavelength using pantograph head acceleration
  measurements,'' in \emph{2017 IEEE International Instrumentation and
  Measurement Technology Conference (I2MTC)}, 2017, pp. 1--6.

\bibitem{PREETI2019}
\BIBentryALTinterwordspacing
M.~Preeti, {Koushik Guha}, K.~Baishnab, K.~Dusarlapudi, and K.~{Narasimha
  Raju}, ``Low frequency mems accelerometers in health monitoring – a review
  based on material and design aspects,'' \emph{Materials Today: Proceedings},
  vol.~18, pp. 2152 -- 2157, 2019, 2nd International Conference on Applied
  Sciences and Technology (ICAST-2019): Material Science. [Online]. Available:
  \url{http://www.sciencedirect.com/science/article/pii/S2214785319320310}
\BIBentrySTDinterwordspacing

\bibitem{Jerri1998}
A.~Jerri, \emph{The Gibbs Phenomenon in Fourier Analysis, Splines and Wavelet
  Approximations}.\hskip 1em plus 0.5em minus 0.4em\relax Dordrecht,
  Netherlands: Kluwer Academic Publishers, 1998.

\bibitem{harris1978}
F.~Harris, ``On the use of windows for harmonic analysis with the discrete
  fourier transform,'' \emph{Proceedings of the IEEE}, vol.~66, pp. 55--83,
  1978.

\bibitem{Brigham1988}
E.~O. Brigham, \emph{The Fast Fourier Transform and Its Applications}.\hskip
  1em plus 0.5em minus 0.4em\relax USA: Prentice-Hall, Inc., 1988.

\bibitem{Borden2014}
A.~R. Borden and B.~C. Lesieutre, ``Variable projection method for power system
  modal identification,'' \emph{IEEE Transactions on Power Systems}, vol.~29,
  no.~6, pp. 2613--2620, 2014.

\bibitem{ChenGPS2018}
G.-Y. Chen, M.~Gan, C.~L.~P. Chen, and L.~Chen, ``A two-stage estimation
  algorithm based on variable projection method for gps positioning,''
  \emph{IEEE Transactions on Instrumentation and Measurement}, vol.~67, no.~11,
  pp. 2518--2525, 2018.

\bibitem{Abel1994}
J.~Abel, ``A variable projection method for additive components with
  application to gps,'' \emph{IEEE Transactions on Aerospace and Electronic
  Systems}, vol.~30, no.~3, pp. 928--930, 1994.

\bibitem{Robinson2007}
\BIBentryALTinterwordspacing
D.~Robinson, S.~Farsiu, and P.~Milanfar, ``{Optimal Registration Of Aliased
  Images Using Variable Projection With Applications To Super-Resolution},''
  \emph{The Computer Journal}, vol.~52, no.~1, pp. 31--42, 04 2007. [Online].
  Available: \url{https://doi.org/10.1093/comjnl/bxm007}
\BIBentrySTDinterwordspacing

\bibitem{Sheng2008}
J.~Sheng and L.~Ying, ``A variable projection approach to parallel magnetic
  resonance imaging,'' in \emph{2008 5th IEEE International Symposium on
  Biomedical Imaging: From Nano to Macro}, 2008, pp. 1027--1030.

\bibitem{Aravkin2012}
A.~Aravkin, T.~van Leeuwen, and N.~Tu, ``Sparse seismic imaging using variable
  projection,'' in \emph{ICASSP, IEEE International Conference on Acoustics,
  Speech and Signal Processing}, 2012.

\bibitem{Gao2014}
F.~Gao, M.~Li, A.~Abubakar, and T.~M. Habashy, ``Application of variable
  projection scheme for data calibration in electromagnetic wave inversion,''
  in \emph{2014 IEEE Antennas and Propagation Society International Symposium
  (APSURSI)}, 2014, pp. 655--656.

\bibitem{ChenRegularized2019}
G.-Y. Chen, M.~Gan, C.~L.~P. Chen, and H.-X. Li, ``A regularized variable
  projection algorithm for separable nonlinear least-squares problems,''
  \emph{IEEE Transactions on Automatic Control}, vol.~64, no.~2, pp. 526--537,
  2019.

\bibitem{Handler2021}
J.~Handler, D.~Ninevski, and P.~O'Leary, ``Decomposition of a periodic
  perturbed signal with unknown perturbation frequency by the method of
  variable projection,'' in \emph{7th International Conference on Mechanical
  Engineering and Automation Science (ICMEAS 2021)}, (to appear) 2021.

\bibitem{Wahba1990}
\BIBentryALTinterwordspacing
G.~Wahba, \emph{Spline Models for Observational Data}.\hskip 1em plus 0.5em
  minus 0.4em\relax Society for Industrial and Applied Mathematics, 1990.
  [Online]. Available:
  \url{https://epubs.siam.org/doi/abs/10.1137/1.9781611970128}
\BIBentrySTDinterwordspacing

\bibitem{PiegTill1996}
L.~Piegl and W.~Tiller, \emph{The NURBS Book}, 2nd~ed.\hskip 1em plus 0.5em
  minus 0.4em\relax New York, NY, USA: Springer-Verlag, 1996.

\bibitem{FileExchange}
J.~Handler, September, 2021, MATLAB Central File Exchange, [Online].
  https://de.mathworks.com/matlabcentral/profile/authors/23787014.

\bibitem{Brandt1998}
G.~Gowan and S.~Brandt, \emph{Data Analysis: Statistical and Computational
  Methods for Scientists and Engineers}, ser. Ohlin Lectures; 7.\hskip 1em plus
  0.5em minus 0.4em\relax Springer New York, 1998.

\bibitem{Weierstrass1885}
K.~Weierstrass, ``\"uber die analytische darstellbarkeit sogenannter
  willk\"urlicher functionen einer reellen ver\"anderlichen,''
  \emph{Sitzungsberichte der Königlich Preußischen Akademie der
  Wissenschaften zu Berlin, 1885 (II)}, pp. 633--639, 789--805, 1885.

\bibitem{DeBoorSplines}
C.~de~Boor, \emph{A Practical Guide to Spline}, 01 1978, vol. Volume 27.

\bibitem{Dung2017}
V.~T. Dung and T.~Tjahjowidodo, ``A direct method to solve optimal knots of
  b-spline curves: An application for non-uniform b-spline curves fitting,'' in
  \emph{PloS one}, 2017.

\bibitem{Ninevski2020}
D.~Ninevski and P.~O'Leary, ``Detection of derivative discontinuities in
  observational data,'' in \emph{Advances in Intelligent Data Analysis XVIII},
  M.~R. Berthold, A.~Feelders, and G.~Krempl, Eds.\hskip 1em plus 0.5em minus
  0.4em\relax Cham: Springer International Publishing, 2020, pp. 366--378.

\bibitem{DPOLeary2013}
D.~{O'Leary} and B.~{Rust}, ``Variable projection for nonlinear least squares
  problems,'' \emph{Computational Optimization and Applications}, vol.~54,
  no.~3, pp. 579--593, 2013.

\bibitem{Istace1995}
\BIBentryALTinterwordspacing
M.-P. Istace and J.-P. Thiran, ``On the third and fourth zolotarev problems in
  the complex plane,'' \emph{SIAM Journal on Numerical Analysis}, vol.~32,
  no.~1, pp. 249--259, 1995. [Online]. Available:
  \url{http://www.jstor.org/stable/2158295}
\BIBentrySTDinterwordspacing

\bibitem{Friedlander1984}
B.~Friedlander and P.~Boaz, ``The modified yule-walker method of arma spectral
  estimation,'' \emph{IEEE Transactions on Aerospace Electronic Systems}, vol.
  AES-20, no.~2, 1984.

\end{thebibliography}

\end{document}